\documentclass[namedreferences]{kluwer}
\usepackage{epsfig}

\begin{document}
\begin{article}

\begin{opening}

\title{Magnetic Interaction Between Stars And Accretion Disks}
\author{Dmitri A. \surname{Uzdensky}\email{uzdensky@kitp.ucsb.edu}
\thanks{KITP preprint NSF-KITP-03-83}}
\institute{Kavli Inst. for Theoretical Physics, Univ. of California,
Santa Barbara, CA 93106}

\runningtitle{Magnetic Star--Disk Interaction}
\runningauthor{Dmitri A. Uzdensky} 

\begin{abstract}

In this review I consider modern theoretical models of coupled 
star--disk magnetospheres. I discuss a number of models, both 
stationary and time-dependent, and examine what physical conditions 
govern the selection of a preferred model.
\end{abstract}

\end{opening}

\section{Introduction}
\label{sec-intro}

In this paper I review recent theoretical progress in our
understanding of magnetic interaction between Young Stellar 
Objects (YSOs), in particular, Classical T-Tauri Stars (CTTSs), 
and their accretion disks. That such interaction takes place, 
we have no doubt, as there is now ample observational evidence 
of strong (of order $10^3$~G) magnetic fields in these systems 
(e.g., \opencite{JK-1999}; \opencite{Guenther-1999}).

\begin{figure}[t]
\centerline{\includegraphics[width=8cm]{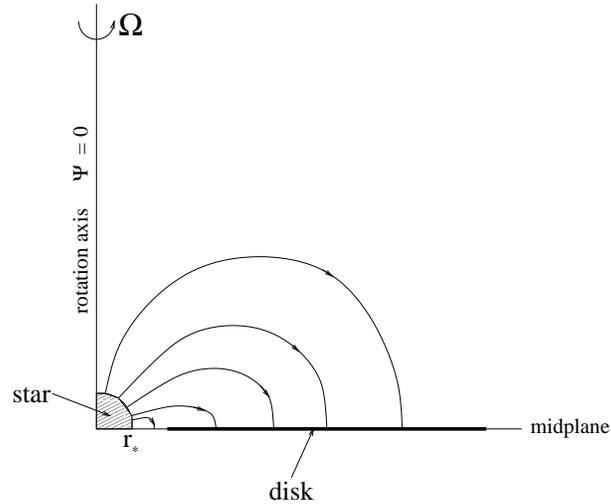}}
\caption{The general geometry of a magnetically-linked 
star--disk system.}
\label{fig-geometry}
\end{figure}

Most of the theoretical work on magnetically linked star--disk systems, 
both analytical and numerical, has focussed on examining the structure 
and role of a large-scale axisymmetric magnetic field with (at least 
initially) dipole-like topology (see Fig.~\ref{fig-geometry}). This 
field presumably arises due to the star's internal magnetic dipole 
moment. Studying such a large-scale star--disk magnetosphere will 
also be the focus of this paper. I will thus ignore the effects 
of any small-scale intermittent loops that may be generated by 
the turbulent dynamo action in the disk.

While I am restricting myself to only large {\it spatial} scales, 
I will consider a variety of {\it temporal} scales. The shortest 
relevant time-scale is the rotation period, typically a few days 
for CTTSs; the longest time-scale is the accretion disk life-time, 
which can be $\sim 10^6$~years or more (\opencite{Konigl-1991}; 
\opencite{Kenyon-1995}).
Among the models developed to date, there exist a dichotomy 
with respect to the system's behavior on the rotation
time-scale. More specifically, in some models a direct magnetic 
connection between the disk and the star is maintained in a
stationary configuration, whereas in other models it is not. 
In this paper I will review both of these classes of models.

Before I proceed, I would like to list some of the most important 
questions related to the subject of this review:\\
1) What physical parameters determine whether a direct star--disk 
coupling via a large-scale dipole-like magnetic field can be maintained 
on the rotation-period time-scale? \\
2) If a quasi-stationary magnetically-coupled configuration does exist, 
what is its structure and how does it evolve on the longer (e.g., accretion) 
time-scale? \\
3) If the magnetic link is disrupted, then what is the non-steady 
process? Are there periodic or quasi-periodic openings and closings
of the field (due to magnetic reconnection) or there is a transition
to a wind-supported permanent stationary open-field configuration 
without the link? \\
4) Is it possible that both scenarios are possible under different
physical circumstances? \\
5) In either scenario, what are the effects of turbulent 
viscosity and magnetic diffusivity? And what is the role 
of winds and jets? \\
6) What are the implications for the time-variability of the
accretion flow and for the angular momentum and energy exchange?
What are the observational consequences that would allow one to 
discriminate between the models?

Although I will not be able to answer all of these questions in this 
review, I will use them as the main guiding themes in my discussion.


\section{Non-stationary models}
\label{sec-non-stationary}

The study of magnetic interaction between YSOs and their disks has 
been pioneered by \inlinecite{Konigl-1991}. He has successfully applied 
the model of \inlinecite{Ghosh-1978}, developed originally for accreting
neutron stars, to explain several important observational features 
of T-Tauri stars, such as their relatively long rotation periods, 
UV-excesses, and inverse P-Cygni profiles. In this steady-state 
model, the stellar magnetic field penetrates the disk over a finite range 
of radii both inside and outside of the corotation radius $r_{\rm co}\equiv
(GM_*/\Omega_*^2)^{1/3}$, where $M_*$ is the mass of the star and $\Omega_*$ 
is its angular velocity. The spin-up magnetic torque due to the field 
lines connecting to the disk inside $r_{\rm co}$ is balanced by the 
spin-down torque by the lines connecting to the disk outside $r_{\rm co}$. 
K{\"o}nigl has proposed that a magnetic field of $10^3$~G at the stellar 
surface (a value consistent with observations, see \opencite{Bouvier-2003}) 
can disrupt the disk at a few stellar radii (but well inside the corotation 
radius) and channel the accretion flow to higher stellar latitudes. 
He also has estimated the typical time needed to bring the star into 
the spin equilibrium with the disk to be $\sim 10^5$ years, much shorter 
than the typical accretion time for CTTSs. 

Although this model has been very successful in explaining many spectral 
and variability features, it has not considered the dynamics of the 
magnetic field itself. The presence of a strong magnetic field has 
just been inferred, but no equations governing the magnetosphere 
structure have been solved. It turns out that there exists a very 
robust mechanism that leads to the breaking of the magnetic link 
on the rotation-period time-scale. This presents a serious obstacle 
for all steady-state models and thus gives us the motivation to consider 
nonstationary models.

The basic idea can be explained as follows. Both the star and the disk 
are fairly good conductors and so the magnetic field can generally be 
considered frozen into them. In addition, they rotate with different 
angular velocities (except at $r_{\rm co}$). Therefore, the field lines 
are twisted by the differential rotation and toroidal magnetic flux is 
generated out of the poloidal flux. As the toroidal field builds up, 
the corresponding field pressure tends to push the field lines outward
and inflate them. At first, the poloidal field structure changes very 
little, but, after the relative star-disk twist angle exceeds one radian 
or so, the field lines start to expand faster and faster (at an angle of 
$\sim 60^\circ$ with respect to the rotation axis) and tend to open up, 
thereby destroying the magnetic link between the star and the disk.

This opening process is essentially identical to a similar
process that have been studied extensively in solar-corona
context (e.g., \opencite{Aly-1984}, \citeyear{Aly-1995}; 
\opencite{Low-1977}, \citeyear{Low-2001}; \opencite{Mikic-1994}; 
\opencite{Uzdensky-2002b}). 
Indeed, the opening of coronal magnetic arcades,
brought about by displacements of the field-line
footpoints on the photosphere, is one of the leading 
mechanisms for Coronal Mass Ejections \cite{Low-2001}. 
A significant amount of work on this process has also been done in 
the accretion-disk context. Thus, in the force-free approximation, 
it has been shown, using both simple analytical and semi-analytical 
arguments (\opencite{Aly-1990}; \opencite{vB-1994}; \opencite{LBB-1994}; 
\opencite{UKL-1}; \opencite{Uzdensky-2002a}; \opencite{LB-2003}), and 
via numerical solutions of the force-free Grad--Shafranov equation 
\cite{UKL-1} that such an opening occurs at a finite twist angle 
(see \opencite{Uzdensky-2002b} for a review). In addition, full 
numerical 2D MHD simulations (without the force-free assumption) 
have demonstrated the opening process at work as a part of the 
overall cycle (\opencite{Hayashi-1996}, \citeyear{Hayashi-2000}; 
\opencite{Goodson-1997}, \citeyear{Goodson-1999a}; \opencite
{Goodson-1999b}; \opencite{Matt-2002}). Thus, at present there 
is no doubt that, if stellar dipole magnetic field penetrates a 
conducting disk over a wide range of radii, the twisting of the 
field lines will open them, thereby breaking the star--disk 
connection everywhere, with perhaps the exception of the inner 
disk region where the magnetic field is strong enough to make 
the matter corotate with the star. This process naturally results 
in a non-steady behavior, which has lead to the development of a 
number of time-dependent models. In addition, however, there exist 
several alternatives, leading to a small number of distinct 
stationary models. I shall discuss the non-steady models first.

After the field lines expand and effectively open 
up, a natural question to ask is: what happens next?
Currently, the situation is not entirely clear and 
there is no unique answer to this question. There are 
two drastically-different possibilities that are most 
often discussed.

\begin{figure}[t]
\centerline{\includegraphics[width=8cm]{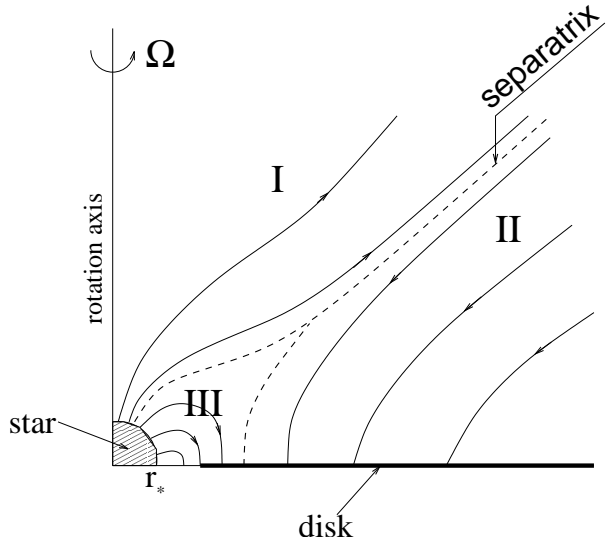}}
\caption{The magnetic configuration of Lovelace et al. (1995).}
\label{fig-LRBK}
\end{figure}

In the first scenario \cite{Lovelace-1995}, developed in the neutron 
star context, once the field lines open, they stay open indefinitely. 
A steady state is then achieved, although it is very different from 
the original one, as the magnetic link between the disk and the star 
has been severed on most of the field lines (see Fig.~\ref{fig-LRBK}). 
One can identify three topologically-distinct regions in the magnetosphere: 
the stellar wind region (region~I), where the field lines extend from the 
star to infinity, the disk wind region (region~II), where the field lines 
return from infinity to the disk, and the closed field region (region~III) 
--- the remnant of the linked magnetosphere, where the field enforces 
corotation of the disk with the star. Thus, the configuration here is 
stationary, but the magnetic link extends only over a small part of 
the disk.

In the second scenario, the situation is really time-dependent,
with quasi-periodic cycles of field inflation and opening due to 
twisting followed by the field closing through reconnection and 
subsequent contraction back to the initial state. This picture
has been suggested by \inlinecite{vB-1994} and has subsequently 
been studied in extensive numerical simulations by a number of
authors (\opencite{Goodson-1997}, \citeyear{Goodson-1999a};
\opencite{Goodson-1999b}; \opencite{Hayashi-2000}; \opencite{Matt-2002}). 
Some recent observational results also seem to favor this point 
of view (e.g., \opencite{Bouvier-2003}). Let us consider this 
most interesting scenario in more detail.

First, notice that in the \inlinecite{Lovelace-1995} scenario the poloidal 
magnetic field reverses across the separatrix between regions I and~II. 
This makes the separatrix an obvious prospective site for reconnection. 
Indeed, the presence of a rather large anomalous or numerical resistivity 
has routinely lead to reconnection in the numerical simulations by 
\citeauthor{Hayashi-1996} (\citeyear{Hayashi-1996}, \citeyear{Hayashi-2000}) 
and by \citeauthor{Goodson-1997} (\citeyear{Goodson-1997}, 
\citeyear{Goodson-1999a}) and \inlinecite{Matt-2002}, .%
\footnote
{We also have to mention that \inlinecite{UKL-2} were sceptical about 
the possibility of reconnection, but this is because they had used the 
\inlinecite{vB-1994} self-similar model describing a uniformly-rotating 
disk. In that model finite-time field-opening occurs without current-sheet 
formation along the separatrix. In a more realistic case of non-uniform 
rotation, \inlinecite{Uzdensky-2002a} has argued that there will be 
finite-time partial field opening accompanied by asymptotic thinning 
of the separatrix current-concentration region, which can be regarded 
as current-sheet formation. As the current layer becomes thinner and 
thinner, a reconnection process may be triggered by anomalous resistivity 
or the Hall effect.}

It is also important to realize that most of the toroidal flux, 
generated in the twisting process, has now been evacuated radially 
to infinity (since the toroidal flux on an inflated flux tube is 
concentrated near the tube's apex). 
Therefore, magnetic field in both open-field regions is essentially 
poloidal (it is exactly poloidal in the force-free framework, but 
some toroidal field may be present in the MHD-wind regime where 
matter inertia is important). This means that if magnetic reconnection 
does occur somewhere along the separatrix, the inner newly-reconnected 
field lines (connected to both the star and the disk) find themselves
out of force-free balance: they have very strong poloidal-field tension 
that tries to pull them back towards the star but almost no toroidal-field 
pressure. As a result, in the absence of a powerful outgoing wind (see the 
discussion below), these inner reconnected field lines contract on the 
alfv{\'e}nic time-scale. If both reconnection (exhibited as a flare) and 
the subsequent contraction and relaxation occur quickly enough, then
the resulting closed field lines have very little residual twist, 
similar to the original dipole-like state. This sets the stage for 
a new cycle. The continuing differential rotation gradually twists 
the lines up again and the whole sequence of events repeats, with 
the natural period of the order of the rotation period. As for the other, 
outer, newly-reconnected field lines, they, together with the apex regions 
of the field lines that are still expanding  somewhere far away, form a 
toroidal plasmoid (in a sense, a flying spheromak). The closed magnetic 
flux surfaces comprising such a plasmoid have the shape of tori nested 
around a circular line (an O-point in poloidal projection). Each plasmoid 
is not magnetically connected to either the disk or the star and is out of 
equilibrium; it then just flies away. If the motion of these plasmoids is 
collimated towards the axis (i.e., if they are flying mostly vertically), 
then they can feed the jet, providing an explanation for the observed 
knotty jet structure \cite{Goodson-1999a}. The plasmoids will be ejected 
out with the time intervals equal to the opening/closing period. For CTTSs, 
however, this period is expected to be too short compared with the observed 
interval between jet knots \cite{Goodson-1999a}.

Whereas the effective field-opening time is about a fraction of 
the rotation period, the time between opening and reconnection is 
not certain. It depends on the intricate details of the reconnection
process and, in particular, is intimately related to the so-called 
reconnection-trigger, or sudden-onset, problem, very well known in 
studies of flares in solar physics \cite{Priest-1984}. Here is what it 
means in the context of our problem. As the field lines start to open, 
one by one, and the current sheet is formed along the separatrix, how 
long does one wait before reconnection starts? In other words, how much 
flux is opened before it is reconnected back? For example, one can imagine 
that reconnection is triggered as soon as the first few field lines have 
opened; then one will see the ejection of small islands, separated 
by the time it takes the critical amount of flux to open (much shorter 
than the rotation period). Or, in the opposite extreme, it may be 
that a large portion or all of the flux opens and only long after 
that reconnection somehow starts; then one will see finite-size 
plasmoids ejected with the time interval equal to the sum of the 
opening time (days) and the uncertain time delay before reconnection 
onset [for example, in the simulations by \inlinecite{Goodson-1999a} 
the total cycle period was about a month].

Also not clear is how much flux is reconnected in each event before 
the reconnection process shuts off. This question is important because 
it determines the size of the ejected plasmoids. Indeed, it may be that 
reconnection proceeds until a large fraction of the flux is reconnected, 
in which case there will be large-amplitude oscillations in all of the 
system's parameters (\opencite{Hayashi-1996}, \citeyear{Hayashi-2000}; 
\opencite{Goodson-1999a}). Alternatively, reconnection may
stop very quickly after it has begun, and then one will see very small 
plasmoids ejected but the larger-scale open-field structure will stay 
intact, as seen in numerical simulations by \inlinecite{Fendt-2002}. 
In fact, the model of \inlinecite{Lovelace-1995} can be viewed as an 
extreme manifestation of this latter scenario. Indeed, in this model 
it has been assumed, without much discussion or argumentation, that 
there is no reconnection at all. One can actually bring forth some 
arguments in favor of this point of view. 

In general, there is a competition between field-line closing via 
reconnection and field-line opening by the wind. Reconnection will 
be stopped if the wind flowing along the open field lines is so strong 
that any newly-reconnected field lines are swept open by it (B.C. Low, 
private communication). More specifically, if there were no flows along 
the unreconnected field lines (i.e., no wind), then a newly reconnected 
closed field line on the inner side of the reconnection region would 
contract rapidly, with the field-line apex moving out of the reconnection 
region towards the star with the poloidal alfv{\'e}n velocity 
$V_{A,\rm pol}$. However, if there is a background outflow such 
as a wind, then one has to add the velocity of this outflow. If 
the latter is larger than $V_{A,\rm pol}$, then the resulting apex 
motion will be directed outward, i.e., the field line will open again.
Thus, I suggest the following physical criterion for determining when 
the re-closing of the open field lines via reconnection will occur.
I propose that if the prospective reconnection site is located 
outside of the Alfv{\'e}n radius (along the separatrix field line), 
so that the wind there is super-alfv{\'e}nic with respect to the 
reconnecting poloidal field, then everything will be swept outward 
by the wind and hence reconnection will not take place and the magnetic 
link will not be re-established. One then will get a helmet-streamer 
configuration like that of \inlinecite{Lovelace-1995}. In the opposite 
case, reconnection will occur and will lead to the closing of (a portion 
of) the field lines, leading to a cyclic behavior. 

To summarize, the system's behavior depends on both the reconnection 
physics (one needs to know where and when reconnection will be triggered) 
and on the wind physics: one needs to have a model for the wind to 
determine everything self-consistently.


\section{Steady-state models}
\label{sec-stationary}

I shall now switch gears and discuss the few existing steady-state 
models in which the magnetic link between the star and the disk remains 
unbroken. First, to maintain the link, one must find a mechanism that 
could stop the twisting process. One obvious possibility is the toroidal 
resistive slippage of the field lines with respect to the plasma in the 
disk. Let us examine this possibility in more detail and see under what 
conditions it can work.

The situation depends critically on the disk's effective magnetic 
diffusivity (which we shall sometimes call the resistivity). 
Unfortunately, the value of this diffusivity is not very well 
known (e.g., \opencite{Bouvier-2003}). Nevertheless, one can set 
a reasonable upper limit by assuming that it is caused by the same 
turbulence that facilitates angular momentum transport across the 
disk. Thus, one can set the magnetic diffusivity~$\eta$ to be equal 
to the Shakura--Sunyaev kinematic viscosity: $\eta\sim \nu_{\rm turb} 
=\alpha c_s h$, where $c_s$ is the speed of sound, $h$ is the disk 
half-thickness, and $\alpha \simeq 0.01-0.1$. This range of the 
$\alpha$-values is consistent with the results of numerical MHD 
simulations of the Magneto-Rotational Instability (e.g., \opencite
{Brandenburg-1996}; \opencite{Stone-1996}). It is also consistent 
with the level of MHD turbulence that is necessary for the ejection 
of disk winds, as follows from the work of \inlinecite{Ferreira-1997} 
combined with the results of \inlinecite{Ferro-Fontan-2003}.%
\footnote
{The conditions for launching MHD winds from accretion disks 
are found to be close to those necessary for the operation of
the Magneto-Rotational Instability \cite{Ferro-Fontan-2003}.}
The effective magnetic diffusivity of this kind leads to the toroidal 
slippage of field lines with respect to the disk with the relative 
drift velocity 
$\Delta v_{\phi}=(\eta/h)|B_\phi/B_z|_d\sim\alpha c_s|B_\phi/B_z|_d$, 
where the subscript $d$ designates the disk's surface. For a thin disk, 
$c_s/v_K \sim h/r \ll 1$; thus, the slippage velocity is usually much 
smaller than the differential rotation velocity $r\Delta\Omega(r)\equiv 
r[\Omega_K(r)-\Omega_*]$. There are, however, two special circumstances 
when this is not so. They are very important as they point us toward the 
ways to get a steady-state configuration. I would like to stress, however, 
that both of these circumstances are somewhat unusual and hence the 
resulting steady states are not very natural. The first of the two 
schemes is realized when the field lines under consideration are 
very close to the corotation radius (so that $\Delta\Omega\ll\Omega_K$); 
it provides the conceptual basis for the model developed by \inlinecite
{Shu-1994a} (see also \opencite{Shu-1994b}; \opencite{Shu-1994c}; 
\opencite{Shu-1995}). The second scheme requires a very large ratio 
$|B_\phi/B_z|_d$ and provides the basis for the model developed by 
\inlinecite{Bardou-1996} and by \inlinecite{Agapitou-2000}. 

\begin{figure}[t]
\centerline{\includegraphics[width=8cm]{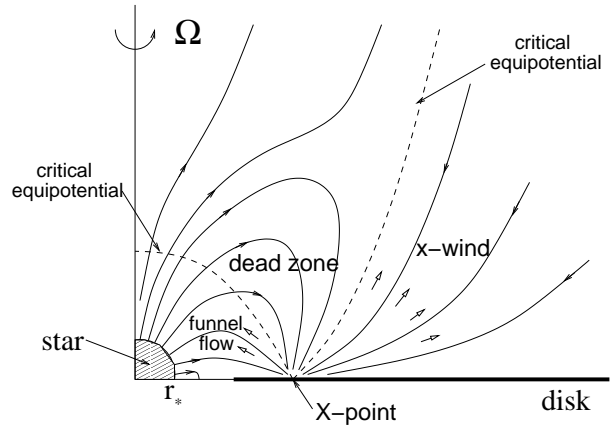}}
\caption{Schematic diagram of the X-wind configuration in 
the Shu et al. (1994) model.}
\label{fig-shu}
\end{figure}

Let us discuss the first scheme first. Usually, its main idea can 
be readily dismissed because for a steady state to exist globally, 
it must exist for all the field lines, and in the majority of models 
most of the magnetic flux crosses the disk a finite distance away from 
$r_{\rm co}$. However, the model developed by \inlinecite{Shu-1994a} 
solves this problem by assuming that almost all of the magnetic flux 
is ``trapped'' and concentrated in the so-called X-region, a very 
close vicinity of the corotation radius (which the authors of the 
model call the X-point). This model is one of the most promising, 
well-developed, and sophisticated models and has gained a lot of 
popularity and observational support (e.g., \opencite{JK-2002}) 
over the last few years. The disk's magnetosphere consists of 
three parts (see Fig.~\ref{fig-shu}). Field lines emanating from 
the inner part of the X-region connect to the star and form the 
magnetic funnel that directs the accretion flow. A second portion 
of the field lines also connects to the star but carries no mass flow; 
it forms what the authors call the dead zone. Finally, the remaining 
field lines, those emanating from the outermost part of the X-region, 
are open and carry the wind that plays a key role in removing the excess 
angular momentum from the disk. In addition, there are some open 
stellar field lines that extend from the star to infinity. Thus, 
the general topology of the poloidal magnetic field is similar to 
the helmet streamer configuration of \inlinecite{Lovelace-1995}. 
An important difference however is that in the \inlinecite{Lovelace-1995} 
model, the poloidal flux is spread smoothly over the entire disk surface 
and the corotation radius plays no special role, whereas in the Shu model
the flux is concentrated close to the X-point, with almost no field at 
$r>r_{\rm co}$. Because of that, the differential rotation in the latter
model is weak ($\Delta\Omega \ll \Omega_*$), and even a small disk 
resistivity is sufficient to eliminate the twisting and hence to 
ensure a steady state. In addition, inside of the corotation radius, 
the disk density drops rapidly as the plasma is uplifted to form the 
funnel flow; as a result, the corotation with the star is enforced by 
the strong magnetic field there.

The inner radius of the disk in the Shu model essentially 
coincides with the X-point [see, however, \inlinecite{Shu-1995},
who put it at $r_{\rm in}=0.74 r_{\rm co}$]. As the accreting 
matter enters the X-region, some part of it is loaded onto the
open field lines and forms the outgoing X-wind. The rest of 
the plasma gradually diffuses through the magnetic field in 
the X-region and is loaded onto the funnel-region field lines
and then falls onto the star. At the same time, most of the 
angular momentum of this falling matter is taken away by the 
magnetic field and is transported back to the inner portion 
of the disk, while only a small fraction ends up on the star. 
This provides an effective control mechanism for the star's spin 
and suggests a plausible explanation of the relatively-long rotation 
periods of CTTSs \cite{Shu-1994a}.

This very interesting model is not free of its own problems and
inconsistencies, however. Thus, for example, it is highly unlikely
that the poloidal field on the inner side of the X-region will not
diffuse towards the star. Indeed, these field lines are very strongly 
bent so that there is a highly concentrated current, essentially an 
equatorial current sheet, between the X-point and the inner edge of 
the disk [i.e., the kink-point of \inlinecite{Shu-1995}]. Any small 
amount of resistivity will then cause the field to slip inward 
through this current layer. As for the plasma flow, it will not 
be able to counter this diffusion because it is in the same 
direction. Thus, the resistive-MHD Ohm's law immediately tells 
us that this configuration cannot be in a steady state.

The Shu model has also been criticized by \inlinecite{Hartmann-1997}. 
He pointed out that their solution requires very fine tuning, so 
that the inner disk radius (determined by the balance between the 
stellar magnetic field and the accretion flow) is equal to $r_{\rm co}$ 
(determined by $\Omega_*$). Hartmann considers this situation unacceptable, 
citing an example of DR Tau, where accretion rate has been observed to 
change on the time scale far too short for $\Omega_*$ to adjust.
He seems to favor the \inlinecite{vB-1994} viewpoint and concludes 
that ``the entire magnetosphere might be a complicated, time dependent
structure''. He also emphasizes the importance of magnetospheric
reconnection, noting that it ``should lead to substantial heating 
and flare activity'' \cite{Hartmann-1997}.

Now let us consider the second possibility for a steady state. 
In this scenario, the balance between the twisting due to the 
differential rotation and the turbulent resistive slippage is 
made possible by a very large ratio of the toroidal to vertical 
magnetic field at the disk surface, of the order of $\alpha^{-1} 
r/h \gg 1$. Such high values are usually considered to be unlikely. 
Indeed, the angle between the field lines and the disk is determined 
by the entire solution in the magnetosphere and cannot be arbitrary. 
The density of matter above the disk is typically so low that magnetic 
forces completely dominate the dynamics there. In this force-free regime, 
the toroidal field at the disk surface, $B_{\phi,d}$, increases in 
proportion to the twist at first, but then reaches a maximum and starts 
to decline during the rapid-expansion phase; it goes to zero as the field 
approaches the open state. The maximum value of $B_{\phi,d}$, achievable 
in a force-free magnetosphere, depends sensitively on the way the poloidal 
magnetic flux is distributed across the disk, that is on the function 
$\Psi_d(r)$. Usually, as it turns out, this maximum value is of the 
same order as the vertical magnetic field and hence the minimum angle 
between the disk and the projection of the magnetic field vector on 
the $\theta-\phi$ plane is of order one. In this case, the differential 
rotation cannot produce the required very large values of the disk toroidal 
field. The primary physical reason for this is that most of the toroidal 
magnetic flux, which is being continuously generated by twisting, becomes 
concentrated near the field-line apex (i.e., the farthest from the star 
point on a field line). As the field expands, it becomes energetically 
favorable for the toroidal flux to escape to infinity by opening the 
poloidal field lines. Coming back to the question of the effects of 
the disk resistivity, we see that, with the disk toroidal field limited 
by the opening and flux-escape process, the toroidal resistive slippage, 
even in a turbulent disk, cannot be fast enough to significantly affect 
the twisting process. 

On the other hand, for a certain class of functions $\Psi_d(r)$,
the maximum value $|B_\phi/B_z|_{\rm d, max}$ of the ratio of the 
toroidal to vertical field components at the disk surface, allowed
by the force-free solution in the magnetosphere, can be large. 
In particular, if $\Psi_d(r)\sim r^{-n}$, then $|B_\phi/B_z|_{\rm d,max} 
\sim O(1/n)$ in the limit $n\rightarrow 0$ (\opencite{LBB-1994}; 
\opencite{Bardou-1996}; \opencite{Agapitou-2000}; \opencite{UKL-1}).

One can then picture the following evolutionary scenario. 
Let us start with a non-steady cyclic configuration such 
as that described by \inlinecite{Goodson-1997}, \shortcite{Goodson-1999a}. 
During the first part of the cycle, as the field expands and approaches 
the open state, the field lines at the disk surface are inclined away 
from the star [i.e., $(B_r/B_z)_d>0$] and hence diffuse a little bit 
outward.%
\footnote{with the initial velocity of order the rms turbulent 
velocity in the disk, $v_{\rm turb}\sim \alpha c_s$; hence 
the characteristic radial footpoint displacement over a rotation
period scales as $\Delta r\sim v_{fp}\Delta\Omega^{-1} \sim
\alpha c_s/\Delta\Omega  \sim \alpha h \Omega_K(r)/
\Delta\Omega \ll r$.}
Then, during the second part of the cycle, as the reconnected 
field contracts back to the nearly potential state, the field 
lines may be inclined towards the star at the disk surface [i.e., 
$(B_r/B_z)_d<0$]; they will then diffuse inward. (Note also that, 
for the field lines inside $r_{\rm co}$, such an inclination is 
conducive to loading of matter onto the field lines; thus, 
accretion can take place during this phase of the cycle.)
Then, one can ask what happens on a much longer time scale, 
when the radial diffusion of magnetic field has to be included. 
Here, two possibilities immediately come to mind.

It may be, as suggested by \inlinecite{vB-1994}, that the little 
diffusive displacements will, over time, redistribute the disk's 
magnetic flux so that the net displacement over one cycle will 
become zero. Then, an averaged steady state will be established, 
i.e., the cycles of field opening, reconnection, and closing will 
produce no net secular evolution in the magnetic flux distribution. 
Since the amount of the outward radial displacement depends on the 
exact moment of reconnection, it follows that the physics of 
reconnection again plays a crucial role in determining the 
long-term magnetic flux distribution.

On the other hand, as I have discussed above, if $\Psi_d(r)\sim r^{-n}$, 
then $|B_\phi/B_z|_{\rm d,max} \sim O(1/n)$ in the limit $n\rightarrow 0$. 
Thus, it is in principle possible for the system to achieve exact, not 
time-averaged, steady state if the disk's flux redistributes in such a 
way that the corresponding value of $n\equiv -d\ln \Psi_d/d\ln r$ becomes 
very small. Since the value of $|B_\phi/B_z|_d$, necessary for the balance 
between the differential rotation and toroidal resistive slippage, is 
inversely proportional to the disk's effective resistivity, we see that 
in this case the disk flux distribution $\Psi_d(r)$ is essentially 
determined by the resistivity. In the case of Shakura--Sunyaev 
turbulent resistivity, one can obtain the following upper limit:
$ n < C |B_\phi/B_z|_{\rm d, max}^{-1} \sim \eta/rh\Delta\Omega =
O(\alpha h/r) \ll 1$, where $C$ is a finite number. Note also 
that for a steady state to be maintained, one must worry not only 
about the toroidal direction, but also about the radial direction.
This requirement gives not just an upper limit, but in fact determines 
implicitly the entire function $n(\eta)$, or, more generally, the 
dependence $\Psi_d(r)[\eta(r)]$ (\opencite{Bardou-1996}; 
\opencite{Agapitou-2000}). Such a stationary field configuration, 
possible in principle, is very different from the dipole field; 
in particular, it leads to a dramatic decrease in the torque 
between the star and the disk\cite{Agapitou-2000}.

\section{Summary}
\label{sec-summary}

In conclusion, I would like to give the following approximate list of 
the major theoretical approaches to the problem of magnetically-coupled 
star--disk magnetospheres:

1) Very rich non-stationary scenario (\opencite{Aly-1990}; 
\opencite{vB-1994}; \opencite{Hayashi-1996}, \citeyear{Hayashi-2000}; 
\opencite{Goodson-1997}, \citeyear{Goodson-1999a}; \opencite{Goodson-1999b};
\opencite{UKL-1}, \citeyear{UKL-2}; \opencite{Matt-2002}) 
with cycles of field inflation, opening, reconnection, contraction, 
and accretion. Both accretion and outflows occur intermittently, 
with variability on the differential rotation period (or somewhat 
longer) time-scale. The amplitude of these oscillations (e.g., how 
much poloidal flux is opened and then reconnected in each cycle) 
depends strongly on the physics of reconnection and is not very 
well constrained. For example, the steady-state model of \inlinecite
{Lovelace-1995} can be considered a limiting case where no reconnection 
takes place at all, and thus the oscillation amplitude is zero.

2) The steady-state X-wind model of \inlinecite{Shu-1994a}. 
The model of \inlinecite{Lovelace-1995} can be considered 
a bridge model between the Goodson and Shu models.

3) Finally, a steady-state closed magnetosphere with the poloidal 
vertical magnetic field that threads the disk scaling as 
$B_z(r)\sim r^{-[2+O(\eta)]}$ --- models of \inlinecite{Bardou-1996} 
and of \inlinecite{Agapitou-2000}. These models take into account 
the field's radial diffusion in the disk over a long (compared with 
$\Omega_*^{-1}$) time scale.

At present, it is apparently too early to select one of these 
models as the preferred one based on purely theoretical considerations.
More rigorous theoretical work, in conjunction with more sophisticated 
and thorough numerical simulations and comparison with observations, 
is needed to sort things out.

I am deeply indebted to Arieh K{\"o}nigl and Christof Litwin for 
many, many insightful and productive discussions, and to Ana 
G{\'o}mez de Castro for her very thoughtful comments and useful 
suggestions. I am also very grateful to the organizers of the 
International Workshop on Magnetic Fields and Star Formation 
(Madrid, April 21--25, 2003) for the invitation to write this 
review. This research was supported by the National Science 
Foundation under Grant No.~PHY99-07949.


{}

\end{article}
\end{document}